\documentclass[nofootinbib,twocolumn,aps,prl,superscriptaddress,longbibliography]{revtex4-2}

\usepackage{amsmath,amssymb,bm}
\usepackage{graphicx}
\usepackage{hyperref}
\usepackage{xcolor}

\begin{document}

\title{Global analysis of tensor mode helicity}

\author{Anna C. Tsai}
\email{anna.tsai@mail.utoronto.ca}
\affiliation{Canadian Institute for Theoretical Astrophysics, 60 St. George Street, Toronto, ON M5S 3H8, Canada}
\affiliation{Department of Physics, University of Toronto, 60 St. George Street, Toronto, ON M5S 1A7, Canada}
\affiliation{Dunlap Institute for Astronomy \& Astrophysics, University of Toronto, AB 120-50 St. George Street, Toronto, ON M5S 3H4, Canada}

\author{Dylan L. Jow}
\email{dylanjow@stanford.edu}
\affiliation{Kavli Institute for Particle Astrophysics and Cosmology, Stanford University, 452 Lomita Mall, Stanford, CA 94305-4085, USA}

\author{Ue-Li Pen}
\affiliation{Institute of Astronomy and Astrophysics, Academia Sinica, Astronomy-Mathematics Building, No. 1, Section 4, Roosevelt Road, Taipei 10617, Taiwan}
\affiliation{Canadian Institute for Theoretical Astrophysics, 60 St. George Street, Toronto, ON M5S 3H8, Canada}
\affiliation{Department of Physics, University of Toronto, 60 St. George Street, Toronto, ON M5S 1A7, Canada}
\affiliation{Dunlap Institute for Astronomy \& Astrophysics, University of Toronto, AB 120-50 St. George Street, Toronto, ON M5S 3H4, Canada}
\affiliation{Perimeter Institute for Theoretical Physics, 31 Caroline St. North, Waterloo, ON N2L 2Y5, Canada}
\affiliation{Canadian Institute for Advanced Research, CIFAR Program in Gravitation and Cosmology, MaRS Centre, West Tower, 661 University Ave., Suite 505, Toronto, ON M5G 1M1, Canada}

\begin{abstract}
A detection of helicity in a primordial gravitational-wave background
(GWB) would provide compelling evidence for parity violation in the
early universe and point toward physics beyond the standard picture of
inflation. Since big bang cosmology is not time translation
invariant, we identify that the so-called ``regular'' and
often-neglected  ``irregular'' solutions are independent degrees of freedom that can in principle carry different, even opposite, helicities. We note that this possibility complicates the interpretation of a measured net helicity.
To determine the
sensitivity of various GW detector configurations  to helicity, we
introduce a universal helicity efficiency coefficient 
that provides a unified geometric description of helicity sensitivity
across gravitational-wave detectors which generalizes and improves 
previous estimators. We apply this helicity efficiency coefficient to
estimate the sensitivity of a range of future GW experiments to
helicity across twenty orders of magnitude in frequency assuming a fully
helical background with an amplitude just below the current
constraints. 
\end{abstract}

\maketitle

\section{Introduction}

A detection of a helical gravitational wave background (GWB) would
represent a fundamental shift in standard assumptions about the
symmetries of the universe. A helical GWB would indicate macroscopic
parity  violation in the universe. Mechanisms  for generating a
helical tensor background can broadly be divided into early- and
late-time channels. Early-time mechanisms source GWs during or before
radiation domination and imprint helicity directly on primordial
tensor modes which then evolve passively.  These mechanisms include
parity-violating inflationary scenarios such as  axion inflation
models \cite{Sorbo2011} or cosmological bounces
\cite{Gielen2017}. Late-time  mechanisms, by contrast, can source GWs
well after horizon re-entry. Topological defects such as cosmic
strings or monopoles radiate GWs continuously as they evolve and
chiral defects could imprint a net helicity on the GWB\footnote{More information on topological defects, including effects on the
cosmic microwave background (CMB), can be found in
\cite{Pen1997,Turok1998}.}.  

A largely unexplored freedom in this picture of helical GW generation is that
the two independent solutions to the second-order differential
equation for tensor perturbations (Eq.~\ref{eq:difeq}), often called 
``regular'' and ``irregular'' modes, described in more detail in the next section, need not share the same
helicity. Since the regular and irregular solutions evolve
independently, and can in principle have contributions from different physical mechanisms,
each mode carries its own helicity. For example, a background could be
purely regular-mode helical, purely irregular-mode helical, helical in
both modes with opposite or matching handedness, or helical in neither. This 
decomposition is well defined on subhorizon scales where regular and
irregular solutions are cleanly distinguished by their oscillatory
sine and cosine behavior. The superhorizon regime, where this
distinction becomes less clean and gauge subtleties enter, is an
interesting direction for future work that will not be pursued in this
study. More details about the role of irregular modes in this regime
can be found in \cite{KodwaniTensors2019}.  Superhorizon irregular
tensor modes potentially rule out inflation, requiring an early period
of cosmic contraction \cite{Gielen2017}.

As global helicity detection would be transformative for physics
beyond the standard cosmological model regardless of which generation
mechanism is ultimately responsible, assessing detector sensitivity to
helicity is a broadly useful exercise. In this paper, we introduce a 
universal helicity efficiency coefficient which unifies and improves
previous estimators and provides a simple criterion for detector
sensitivity to global helicity using a single geometric measure
composed of spatial separation and polarization vectors. Using this
helicity sensitivity condition, we place forecasts on helicity
sensitivity across twenty orders of magnitude in frequency.  In
detector configurations with fixed spatial separations between
polarization measurements, we derive sensitivity scaling relations for
the instrument sensitivity range.  For cosmic microwave background
(CMB) measurements of GWs, there is a sensitivity suppression effect
from the approximate flatness of the sky for $l>2$.

\section{Regular and irregular tensor perturbation solutions}
\label{sec:regular-irregular}

Tensor perturbations of the form,
\begin{align}
    h_{ij}(\eta, \textbf{x}) = \sum_{\lambda} \int \frac{d^3 k}{(2\pi)^3} h_{\lambda}(\eta, \textbf{k}) e^{i\textbf{k} \cdot \textbf{x}} \epsilon^{\lambda}_{ij},
\end{align}
where $\lambda$ denotes the GW polarization, are solutions to a second-order differential equation.
Assuming linear order and homogeneous conditions, the strain $h_{\lambda}(\eta, \boldsymbol{k})$ satisfies:
\begin{align}
    \ddot{h}_{\lambda} + 2 \frac{\dot{a}}{a} \dot{h}_{\lambda} + k^2 h_{\lambda} = 0,
    \label{eq:difeq}
\end{align}
where $\dot{h}_{\lambda} \equiv \frac{dh_{\lambda}}{d\eta}$, and $\eta$ is the conformal time. During radiation domination, $\frac{\dot{a}}{a} \propto \frac{1}{\eta}$, and the solutions to the second-order differential equation become:
\begin{align}
    h^{L/R}(\eta, k) &= A^{L/R}(k) \frac{\sin(k\eta)}{k\eta} + B^{L/R}(k) \frac{\cos(k\eta)}{k\eta}, 
\end{align}
The sine and cosine solutions are called the
regular/sine/growing mode and the irregular/cosine/decaying
mode, respectively. ``Decaying'' and ``growing'' refer to the $\eta \ll 1$ superhorizon regime where the irregular mode is approximated by $1/\eta$, and the regular mode is approximately constant (Fig.~1 of Ref.~\cite{KodwaniScalar2019}).  

Although in this work we will not explore early-time helical GWB generation methods, we briefly summarize the literature to illustrate the broad impact of a helical GWB analysis
that may be informed by the universal helicity analysis presented in
this work. In the standard treatment of tensor perturbations generated
during inflation, irregular modes are normalized so that the amplitude
is negligible on subhorizon scales. This is because the irregular
mode amplitude diverges when approaching the singularity at $\eta = 0$
and decays rapidly for $\eta>0$. In addition, if there is a scalar
inflationary period before radiation domination, this will also
suppress the amplitude of the irregular mode. This treatment of the
irregular mode is equivalent to setting $B^{\lambda} =
0$. Ref.~\cite{KodwaniTensors2019} suggests that the marginalization of
irregular modes is a result of a choice of normalization and argues
that irregular modes could persist on subhorizon scales without
necessitating any divergent observables (the GW strain on superhorizon
scales is not an observable in itself). If one chooses to normalize the amplitude
of the irregular mode to be equal to that of the regular mode solution
at the horizon (Fig.~1 in Ref.~\cite{KodwaniScalar2019}),
one could in principle retain a surviving irregular mode from
early-time generation mechanisms. One class of models capable of
producing an observable irregular tensor mode is provided by bouncing
cosmologies \cite{Gielen2017}. Unlike standard inflation, in which the
Universe undergoes a period of accelerated expansion, bouncing models
posit an initial contracting phase followed by a non-singular bounce
into the present expanding Universe. This effect mixes the two
independent solutions, allowing the mode that would normally decay in
an expanding universe to survive with a non-negligible amplitude after
the bounce. 

There are some proposed observational constraints on the existence of
surviving irregular modes from early-time mechanisms. Superhorizon
tensor perturbation modes can produce anisotropies in local patches of
the universe (e.g. Bianchi cosmologies \cite{Pontzen2009}) which manifest as a shear in the expansion.  Although
Ref.~\cite{Saadeh2016} places constraints on regular and irregular tensor
shear modes using Planck data, Ref.~\cite{KodwaniTensors2019} argues that
these constraints depend sensitively on the choice of
normalization. While Ref.~\cite{Saadeh2016} (see Table 2) suggests that
irregular shear modes can be constrained by a factor of between
$10^{3}$ and $10^{4}$ better than regular modes,
Ref.~\cite{KodwaniTensors2019} finds that when normalizing modes to have
the same amplitude at horizon crossing, irregular modes can only be
constrained by a factor of approximately 10 better than regular
modes. Additional constraints arise from Big Bang Nucleosynthesis
(BBN) from the shear bound found in Ref.~\cite{Campanelli2011}
(Eq.~16). However, this bound applies to the total expansion shear
measured in the matter frame, which includes contributions from both
tensor perturbations and the matter velocity field. Therefore, it does
not directly constrain the primordial tensor amplitude without
assumptions about the accompanying matter configuration.

Late-time mechanisms for stochastic GWB generation, such as topological defects \cite{Pen1997,Turok1998}, can produce both regular and irregular tensor modes without the divergent behavior associated with irregular modes generated at early times. Because the two modes are independent solutions of the tensor perturbation equation, they can carry independent helicities. It is therefore possible, in principle, for a source to produce regular and irregular modes with equal amplitudes but opposite helicities. In such a scenario, the net helicity of the total GWB could vanish even though each mode is individually fully helical. A measurement sensitive only to the mean helicity would then classify the background as unpolarized, masking the distinct helicity content of the two modes. This possibility motivates treating the regular and irregular modes as separate helicity degrees of freedom when interpreting measurements of a helical GWB. In Ref.~\cite{Pen1997,Turok1998}, parity symmetry of the stress-energy tensor is assumed. To our knowledge, the implications of parity-violating defect sources for the relative helicity content of the regular and irregular tensor modes have not been studied in a cosmological setting. Exploring such sources could provide a concrete realization of the independent helicity structure considered here.

\section{Helicity efficiency coefficient}

Here, we introduce a geometric criterion that a gravitational wave detector must meet to be helicity sensitive. We will consider a Gaussian GWB of cosmological origin that is homogeneous, isotropic, and which can be left or right circularly
polarized, or unpolarized. Such a background is fully characterized by the two-point correlation functions of Stokes I (total intensity) and Stokes V (helicity).

The metric perturbation of a single monochromatic GW source propagating from the $\hat{n}$-direction, with some frequency $f_0$, is given by
\begin{align}
    h_{ij}(t, x) = h \epsilon_{ij} e^{-2 \pi i f_0 (\hat{n} \cdot {\bf x} + t - t_0)}.
\end{align}
The polarization tensor, $\epsilon_{ij}$, is a normalized linear combination of the $+$ and $\times$ modes:
\begin{align}
    \epsilon^+ = \hat{\phi}_i \hat{\phi}_j - \hat{\theta}_i \hat{\theta}_j, \\
    \epsilon^\times = \hat{\theta}_i \hat{\phi}_j + \hat{\phi}_i \hat{\theta}_j,
\end{align}
where $\hat{\theta}$ and $\hat{\phi}$ are the spherical-coordinate basis vectors at $\hat{n}$. Left- and right-circular polarized modes are linear combinations of $+$ and $\times$ modes: $\epsilon^L_{ij} = (\epsilon^+_{ij} + i \epsilon^\times_{ij}) / \sqrt{2}$, $ \epsilon^R_{ij} = (\epsilon^+_{ij} - i \epsilon^\times_{ij}) / \sqrt{2}$. 

We introduce a polarization unit vector $\boldsymbol{\hat{e}}$ in the plane perpendicular to $\boldsymbol{\hat{n}}$ which points in the direction of maximum perturbation.  Note that for GWs, there are four degenerate polarization vectors $\boldsymbol{\hat{e}}$ representing the same polarization state (rotations of $\pi/2$). For a left-handed GW, $\boldsymbol{\hat{e}}$ rotates clockwise as the GW propagates, viewed from the origin.  Similarly, $\boldsymbol{\hat{e}}$ rotates counterclockwise for a right-handed GW. The measured polarization vector $\boldsymbol{e_u}$ is the projection of $\boldsymbol{\hat{e}}$ onto the detector polarization vector basis. In the case of an interferometer, this is a projection onto an interferometric arm oriented along a vector $\boldsymbol{\hat{u}}$: $\boldsymbol{e_u} \equiv (\boldsymbol{\hat{e}} \cdot \boldsymbol{\hat{u}}) \boldsymbol{\hat{u}}$.

We express the minimum helical sensitivity criterion using two polarization vector measurements, $(\boldsymbol{e_u})_1$ and $(\boldsymbol{e_u})_2$, measured at spatial separation $\boldsymbol{s}$.  In this geometry, the minimum condition is therefore that  $(\boldsymbol{e_u})_1$,  $(\boldsymbol{e_u})_2$, and $\boldsymbol{s}$ cannot be coplanar, or in other words, the pseudo-scalar triple product must not vanish: 
\begin{align}
    H \equiv \boldsymbol{s} \cdot( (\boldsymbol{e_u})_1 \cdot  (\boldsymbol{e_u})_2)( (\boldsymbol{e_u})_1 \times  (\boldsymbol{e_u})_2) \neq 0.
    \label{eq:helicity_condition}
\end{align}

A simple way to understand this geometric condition is to consider a single, circularly polarized GW propagating along the separation vector. If we align the GW so that $\boldsymbol{\hat{e}}$ is parallel to $\boldsymbol{\hat{u}_1}$ at the first polarization measurement, at the second polarization measurement (taken a distance $s$ away from the first), helicity sensitivity requires that one be able to tell if $\boldsymbol{\hat{e}}$ rotated left or right.  If $s=0$, it is impossible to tell if $\boldsymbol{\hat{e}}$ rotated left or right. Similarly, if polarization basis vectors $\boldsymbol{\hat{u}_1}$ and $\boldsymbol{\hat{u}_2}$ are coplanar with $s$, one will be sensitive to a magnitude difference in $(\boldsymbol{e_u})_1$ versus $(\boldsymbol{e_u})_2$, however it is impossible to tell which direction in the plane tangent to $\boldsymbol{s}$ $\boldsymbol{\hat{e}}$ rotated. The triple product is a mathematical construct which vanishes for coplanar vectors and, because of the spin-2 nature of GWs, for fully orthogonal vectors. $H$ is maximized with $45$ degree separation. Thus, the triple product characterizes the degree to which the detector is sensitive to helicity and we refer to it as the helicity efficiency coefficient or the HEC.

An additional helical sensitivity constraint general to any frequency range is that $s$ must be on the order of a gravitational wavelength. This is a requirement for unambiguously measuring the GW phase change across the separation.

\begin{figure*}[t]
    \centering
    \includegraphics[width=\textwidth]{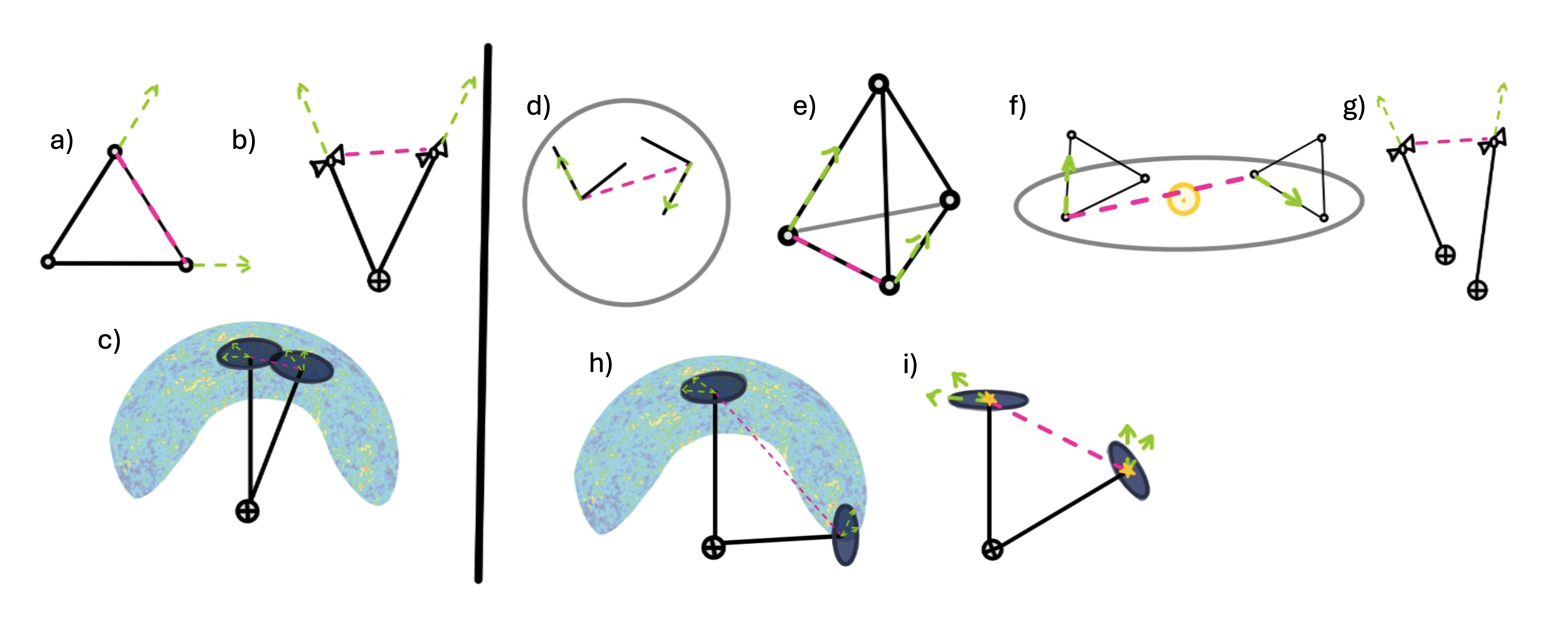}
    \caption{GW detector configurations grouped by helicity-insensitive (left) and helicity sensitive (right). $\boldsymbol{s}$, the separation vector is drawn as a dashed pink line and polarization measurement vectors, $\boldsymbol{e_u}$, are shown as dashed green vectors. a) Proposed LISA configuration composed of three satellites. b) Example two pulsar PTA configuration demonstrating PTA coplanarity. c) CMB measurements at small angular scales measure polarizations nearly coplanar with the separation vector due to local flatness of a sphere, yielding helicity insensitivity. d) LIGO-Livingston-KAGRA, LIGO-Livingston-VIRGO paired configurations are sensitive to helical GWs \cite{Martinovic2021} at $f \sim 30$ Hz, within the most sensitive range of LVK detectors with angular separations on the Earth of $99.9^{\circ}$ and $76.8^{\circ}$ respectively \cite{Chu2021}. e) One additional out-of-plane satellite can be added to LISA for a helicity sensitive space based detector with minimal additions. There are two potential drawbacks to this configuration: the separation $s$ yields helicity sensitivity near $f=10^{-1}$, which is on the edge of the LISA sensitivity band and, the additional satellite will not be in a stable orbit, leading to potential complications caused by varying interferometric arm lengths. f) A two LISA constellation configuration with an approximately $30-40^{\circ}$ angular separation either as a constituent of BBO or on its own \cite{Orlando2021} is helicity sensitive for $f \sim 10^{-3}$ Hz, within the more sensitive range of LISA. g) A (very) hypothetical non-coplanar configuration for PTAs requiring intergalactic collaboration with a second observer $\sim$ a light-year away. h) A CMB $l=2$ search for helicity showing a coplanar detector configuration which still satisfies the conditions of Eq. \ref{eq:helicity_condition}. i) Gaia measurements of stellar angular displacement provide another example of a coplanar detector that is helicity sensitive. }
    \label{fig:HelicityGeometry}
\end{figure*}

In Fig.~\ref{fig:HelicityGeometry}, we group some existing and proposed GW detector configurations into helicity-insensitive (left) and helicity sensitive (right) based on the HEC. When GW detectors measure polarization vectors along the
lines connecting test masses, Eq.~\ref{eq:helicity_condition} reduces to a condition of coplanarity of the geometry of the detector \cite{Smith2017}. This is the case for the LIGO-VIRGO-KAGRA (LVK) ground-based detector configurations (d) as well as the LISA (a) and proposed extensions to LISA (f) configurations. Although PTAs are polarization sensitive to individual point sources, they are not helicity sensitive to an isotropic background (b) as in this case PTAs only measure polarization parallel to the Earth-pulsar baselines. A hypothetical non-coplanar configuration for PTAs would be a system of two spatially separated observers (g).

In cases such as the CMB or astrometry, where polarization vectors are measured in the plane tangent to the line of sight, detector geometry can be coplanar without breaking the helicity sensitivity condition since the polarization vectors span the necessary dimensions. Using CMB EB or TB cross-correlations (c,h), one is sensitive to helicity on the largest angular scales, however on small angular scales, the HEC approaches zero as polarization measurements become approximately coplanar with the separation vector. In the PTA frequency band, astrometric methods are helicity sensitive (i) as polarization is measured in the plane tangent to the line of sight.

\section{Helicity sensitivity of future GW experiments}

In this section, we predict the sensitivity of future GW experiments to a stochastic isotropic, helical GWB using the HEC. Here, we remain agnostic to the strength of any potential GWB; it may be that a cosmological GWB, helical or otherwise, is too weak to be detected in any capacity. Instead, we want to address the question: which methods of detection even have the capability of distinguishing a helical, isotropic GWB of cosmological origin from an unpolarized background? What are the minimum requirements to turn a helicity-insensitive detector into a helicity-sensitive one? We argue that the implications of a global helicity detection for fundamental physics are sufficiently wide-ranging that helicity sensitivity should be a routine consideration in the development of future GW detectors. We first consider the class of detectors with fixed separation, $s$, such as LVK and LISA/future extensions on LISA where we derive simple helical sensitivity scaling relations in frequency. For this class of detectors, polarization is measured along interferometric arms, reducing Eq. \ref{eq:helicity_condition} to a condition of non-coplanarity of the detector geometry itself as described in \cite{Smith2017,Domcke2020}. In this paper, we use these simple frequency scaling relations to derive future helicity sensitivity curves, however more detailed sensitivity calculations can be found in \cite{Smith2017}. We then address astrometric methods, pulsar timing arrays (PTAs), and using the CMB. 

\subsection{GW detectors with fixed separation}
Unlike measuring the total GW power, which can be measured independently by each detector, helicity sensitivity relies on the phase-sensitive cross-correlations between multiple detectors; consequently, rapid inter-detector phase variation causes cancellation of the helicity signal even when the total GW power remains unsuppressed. In this section, we derive the fractional helicity sensitivity suppression as a function of GW frequency compared to the total GW power by calculating angular regions of slow phase variability.

For GW detectors with fixed separation $s$, helicity sensitivity peaks when the gravitational wavelength is on the order of the baseline length, $\lambda_{\rm GW} \sim s$. We motivate the sensitivity drop off when $\lambda_{\rm GW} < s$ and $\lambda_{\rm GW} > s$ by deriving two simple scaling relations with respect to wave number $k$.

The GW phase difference measured at detector one versus detector two (located $\bf s$ away from detector one) is given by 
\begin{align}
    \delta \phi = k s \cos(\theta)
\end{align}
where $k = 2 \pi/ \lambda_{\rm GW}$ and $\theta$ is the angle between $\bf s$ and the direction of GW propagation $-\hat{n}$. The phase difference is considered measurable when $\delta \phi \sim 1$ radian. In the short wavelength case, $\lambda_{\rm GW} < s$, $ks\gg1$ and the region of constructive signal is a thin ring around the baseline $\bf s$, centered at $\theta = \pi/2$. That is to say that waves propagating within a certain range of angles have a phase difference at the two detectors less than a radian and so add constructively, whereas waves outside that range have large phase differences and therefore add as noise. We calculate $\theta_c = \frac{\pi}{2} - \frac{1}{ks}$ using a small angle expansion around $\pi/2$. This means the thickness of the ring of coherence goes as $\frac{2}{ks}$ and the area of the ring scales as $\sim \frac{1}{ks}$. This means that helicity sensitivity will be suppressed by a factor of $\frac{1}{ks}$ in the regime $\lambda_{\rm GW} < s$ compared to a parity symmetric search. We note that in the case of LVK, this suppression factor is already accounted for in the Stokes I sensitivity curve as it is built from a cross-correlation of spatially separated detectors. Therefore the expected Stokes V sensitivity matches the expected Stokes I sensitivity in the high frequency regime.

When $\lambda_{\rm GW} > s$, the detectors respond only to the quadrupolar (spin-2) part of the wave, so the quantity of interest is the correlation between the quadrupolar responses of detectors~1 and~2. Four unit vectors enter this calculation: $\hat{\mathbf{u}}_1$ and $\hat{\mathbf{u}}_2$, the polarization basis vectors of the two detectors, and $\hat{\mathbf{e}}_1$ and $\hat{\mathbf{e}}_2$, the intrinsic GW polarization vector evaluated at the two detectors. We project all four onto the plane perpendicular to the detector separation vector $\mathbf{s}$. In this plane, the projections of $\hat{\mathbf{e}}_1$ and $\hat{\mathbf{e}}_2$ are separated by an angle $\psi = \pm\,\delta\phi/2$, where the upper (lower) sign corresponds to left (right) sources, while the projections of $\hat{\mathbf{u}}_1$ and $\hat{\mathbf{u}}_2$ are separated by an angle $\beta$ set by the detector geometry. If we choose the GW phase so that the projection of $\hat{\mathbf{e}}_1$ is parallel to $\hat{\mathbf{u}}_1$, the correlation between the detectors is proportional to the projection of $\hat{\mathbf{e}}_2$ onto $\hat{\mathbf{u}}_2$, which for a quadrupolar response is $\cos 2(\beta \mp \psi)$ for left and right sources, respectively. The Stokes~$V$ measurement depends on the difference between the cross-correlations of the two GW polarizations, $\cos 2(\beta-\psi) - \cos 2(\beta+\psi) = 2\sin 2\beta\,\sin\delta\phi$, which for small $\delta\phi$ scales linearly with $ks$.


\subsection{GW with variable separation}
While PTAs can measure polarization in the plane tangent to the Earth-pulsar baseline for discrete astrophysical sources, PTAs are not sensitive to a helical stochastic background \cite{Belgacem2020} because in the regime of an unresolvable background, PTAs can only measure polarization along the Earth-pulsar baseline (see Fig. \ref{fig:HelicityGeometry} b). In principle, a way of breaking coplanarity would be to have simultaneous GW detection at two spatially separated observers (i.e. at the Earth and some other location); however, the conditions on GW phase sensitivity (the crux of measuring helicity) require the second observer to be $\sim \rm lightyears$ away, requiring simultaneous measurements at the Earth and a planet orbiting a nearby star for example. 

\begin{figure}[htbp]
    \centering
    \includegraphics[width=1\linewidth]{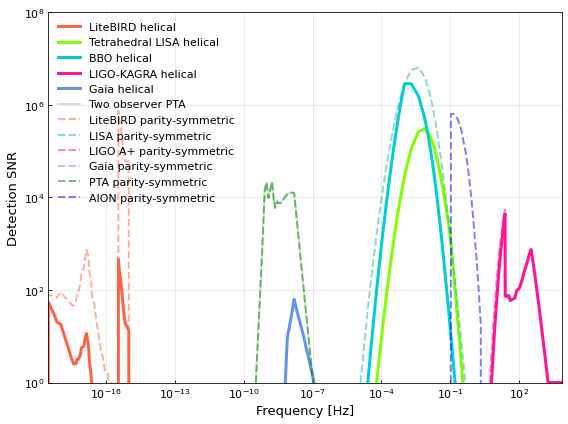}
    \caption{Here we assume that a fully helical isotropic stochastic GWB has a strain just below current detector sensitivity at each frequency. Stokes I detection SNR for future detectors is then the assumed background strain divided by the strain sensitivity of the next generation of detectors (dashed lines marked ``parity-symmetric''). Using solid lines, we show sensitivity to the Stokes V parameter using helicity-sensitive GW detectors with a single separation baseline. For detector configurations with a fixed separation (tetrahedral LISA, BBO, LVK), helicity sensitivity for gravitational wavelengths shorter than $\boldsymbol{s}$ is suppressed by a factor of $1/(ks)$ due to the scaling of the coherence angle. Because the projected future LVK strain sensitivity is already a cross-correlation between spatially separated detectors, the suppression of $1/(ks)$ is already included in the Stokes I curve. Helicity sensitivity for gravitational wavelengths longer than $s$ is suppressed by a factor of $ks$.  The CMB helicity search is suppressed by a factor of $2/l$ following Ref.~\cite{Masui2017}. The projected LiteBIRD sensitivity curve is taken from Ref.~\cite{Campeti2021}, and the projected Gaia sensitivity is taken from Ref.~\cite{Moore2017} and adjusted to be a factor of two more sensitive to account for the 10 year mission duration instead of the 5 year expected duration. For all other experiments, we use sensitivity curves from Ref.~\cite{Renzini2022}.}
    \label{fig:detectionSNR}
\end{figure}

However, in the $10^{-10}-10^{-6}$ Hz frequency range there is still some hope for helicity sensitivity: astrometric methods \cite{Darling2018} are helicity sensitive \cite{Qin2017} as they measure polarization in the plane tangent to the line of sight. Similarly to the CMB, astrometric data can be decomposed into E and B modes \cite{Dai2012, Qin2017}. Non-zero EB correlations would be a tell-tale sign of parity violation. Time-series astrometric data for GW searches is scheduled for release in December 2026 as a part of Gaia DR4 \cite{GaiaDataReleases}, allowing for the possibility of a helical, isotropic GWB search. In Fig. \ref{fig:detectionSNR}, we use the expected detection sensitivity for the planned 5-year mission \cite{Moore2017} and divide by a factor of 2 to represent the estimated sensitivity curve for the extended 10-year observing duration.

In the case of the CMB, for large spherical harmonic $l-$ modes, $s$, $e_{u1}$, and $e_{u2}$  approach coplanarity, as GW polarization from the CMB is measured in the plane tangent to the surface of last scattering. This approximate two dimensionality effect produces a factor of $2/l$ suppression in sensitivity as discussed in Ref.~\cite{Masui2017}.

There are a few proposed experiments that aim to bridge the frequency gaps between ground-based detectors, LISA, and PTA, though they do not show much promise in being helicity sensitive. Lunar Laser Ranging (LLR) \cite{Vanvlasselaer2026} and pulsar binary resonance \cite{Hui2013} are sensitive to GWs in the range of $10^{-7}-10^{-4}$ Hz, though neither are currently more constraining than the $\rm N_{eff}$ \cite{Renzini2022} constraint. Terrestrial atom interferometers \cite{2009PhLB..678...37D} could in principle be built in polarization sensitive configurations, however, constrained to the face of the Earth, helicity sensitivity is maximized at approximately $50$ Hz which is outside the sensitivity range of ground based atom interferometers.

\section{Conclusion}
We present GW detector helicity sensitivity as a single geometric condition: non-coplanarity of the separation vector and the two measured polarization vectors. The helicity efficiency coefficient (HEC) unifies existing detector-specific criteria and explains, in one framework, why pulsar timing arrays are fundamentally blind to helicity, while CMB and astrometric measurements remain sensitive even with apparent coplanar geometry. Applying this coefficient across twenty orders of magnitude in frequency, we find that fixed-baseline detectors peak in helicity sensitivity near $\lambda_{\rm GW}\sim s$, falling off as $1/(ks)$ for shorter wavelengths and $ks$ for longer ones. This points to near-term opportunities requiring only modest changes to planned instruments: a six-satellite LISA extension and paired LIGO-KAGRA/LIGO-Virgo baselines are both helicity sensitive within their most sensitive bands. In the PTA band, where pulsar timing itself is helicity blind, astrometric measurements offer a viable alternative: time-series astrometric data due for release in December 2026 will, for the first time, permit a helicity search in the $10^{-10}-10^{-6}$ Hz range. In addition, we identify a largely unexplored freedom in tensor-mode generation: the regular and irregular solutions to the perturbation equation carry independent, potentially opposite, helicities. A helicity detection---or a substantially tightened upper limit---would therefore probe the physics of the early universe in a way that is complementary to, and independent of, existing constraints on tensor amplitude and spectral shape.

\section*{Acknowledgments}

We thank Neil Turok, Daniel Baumann, Deyan Mihaylov, Kinwang Ng and Kiyo Masui for helpful discussions. U.P. is supported by the Natural Sciences and Engineering Research Council of Canada (NSERC) [funding reference number RGPIN-2019-06770, ALLRP 586559-23, RGPIN-2025-06396], Ontario Research Fund (ORF-RE Fund), and AMD AI Quantum Astro. We used generative AI tools for proofreading this manuscript. The authors reviewed all AI-assisted edits and take full responsibility for the final manuscript.

\bibliographystyle{apsrev4-2}
\bibliography{references}  

@ARTICLE{Smith2017,
       author = {{Smith}, Tristan L. and {Caldwell}, Robert},
        title = "{Sensitivity to a frequency-dependent circular polarization in an isotropic stochastic gravitational wave background}",
      journal = {\prd},
         year = 2017,
        month = feb,
       volume = {95},
       number = {4},
          eid = {044036},
        pages = {044036},
          doi = {10.1103/PhysRevD.95.044036},
archivePrefix = {arXiv},
       eprint = {1609.05901},
 primaryClass = {gr-qc},
       adsurl = {https://ui.adsabs.harvard.edu/abs/2017PhRvD..95d4036S}
}

@ARTICLE{Masui2017,
       author = {{Masui}, Kiyoshi Wesley and {Pen}, Ue-Li and {Turok}, Neil},
        title = "{Two- and Three-Dimensional Probes of Parity in Primordial Gravity Waves}",
      journal = {\prl},
         year = 2017,
        month = jun,
       volume = {118},
       number = {22},
          eid = {221301},
        pages = {221301},
          doi = {10.1103/PhysRevLett.118.221301},
archivePrefix = {arXiv},
       eprint = {1702.06552},
 primaryClass = {astro-ph.CO},
       adsurl = {https://ui.adsabs.harvard.edu/abs/2017PhRvL.118v1301M}
}

@ARTICLE{KodwaniTensors2019,
       author = {{Kodwani}, Darsh and {Meerburg}, P. Daniel and {Pen}, Ue-Li and {Wang}, Xin},
        title = "{Initial conditions of the universe: Decaying tensor modes}",
      journal = {arXiv e-prints},
         year = 2019,
        month = oct,
          eid = {arXiv:1910.01416},
        pages = {arXiv:1910.01416},
          doi = {10.48550/arXiv.1910.01416},
archivePrefix = {arXiv},
       eprint = {1910.01416},
 primaryClass = {astro-ph.CO},
       adsurl = {https://ui.adsabs.harvard.edu/abs/2019arXiv191001416K}
}

@ARTICLE{KodwaniScalar2019,
       author = {{Kodwani}, Darsh and {Meerburg}, P. Daniel and {Pen}, Ue-Li and {Wang}, Xin},
        title = "{Initial conditions of the Universe: A sign of the sine mode}",
      journal = {\prd},
         year = 2019,
        month = jun,
       volume = {99},
       number = {12},
          eid = {123518},
        pages = {123518},
          doi = {10.1103/PhysRevD.99.123518},
archivePrefix = {arXiv},
       eprint = {1903.05042},
 primaryClass = {astro-ph.CO},
       adsurl = {https://ui.adsabs.harvard.edu/abs/2019PhRvD..99l3518K}
}

@ARTICLE{Moore2017,
       author = {{Moore}, Christopher J. and {Mihaylov}, Deyan P. and {Lasenby}, Anthony and {Gilmore}, Gerard},
        title = "{Astrometric Search Method for Individually Resolvable Gravitational Wave Sources with Gaia}",
      journal = {\prl},
         year = 2017,
        month = dec,
       volume = {119},
       number = {26},
          eid = {261102},
        pages = {261102},
          doi = {10.1103/PhysRevLett.119.261102},
archivePrefix = {arXiv},
       eprint = {1707.06239},
 primaryClass = {astro-ph.IM},
       adsurl = {https://ui.adsabs.harvard.edu/abs/2017PhRvL.119z1102M}
}

@ARTICLE{Dai2012,
       author = {{Dai}, Liang and {Kamionkowski}, Marc and {Jeong}, Donghui},
        title = "{Total angular momentum waves for scalar, vector, and tensor fields}",
      journal = {\prd},
         year = 2012,
        month = dec,
       volume = {86},
       number = {12},
          eid = {125013},
        pages = {125013},
          doi = {10.1103/PhysRevD.86.125013},
archivePrefix = {arXiv},
       eprint = {1209.0761},
 primaryClass = {astro-ph.CO},
       adsurl = {https://ui.adsabs.harvard.edu/abs/2012PhRvD..86l5013D}
}

@ARTICLE{Qin2017,
       author = {{Qin}, Wenzer and {Boddy}, Kimberly K. and {Kamionkowski}, Marc and {Dai}, Liang},
        title = "{Pulsar-timing arrays, astrometry, and gravitational waves}",
      journal = {\prd},
         year = 2019,
        month = mar,
       volume = {99},
       number = {6},
          eid = {063002},
        pages = {063002},
          doi = {10.1103/PhysRevD.99.063002},
archivePrefix = {arXiv},
       eprint = {1810.02369},
 primaryClass = {astro-ph.CO},
       adsurl = {https://ui.adsabs.harvard.edu/abs/2019PhRvD..99f3002Q}
}

@ARTICLE{Gielen2017,
       author = {{Gielen}, Steffen and {Turok}, Neil},
        title = "{Quantum propagation across cosmological singularities}",
      journal = {\prd},
         year = 2017,
        month = may,
       volume = {95},
       number = {10},
          eid = {103510},
        pages = {103510},
          doi = {10.1103/PhysRevD.95.103510},
archivePrefix = {arXiv},
       eprint = {1612.02792},
 primaryClass = {gr-qc},
       adsurl = {https://ui.adsabs.harvard.edu/abs/2017PhRvD..95j3510G}
}

@ARTICLE{Saadeh2016,
       author = {{Saadeh}, Daniela and {Feeney}, Stephen M. and {Pontzen}, Andrew and {Peiris}, Hiranya V. and {McEwen}, Jason D.},
        title = "{How Isotropic is the Universe?}",
      journal = {\prl},
         year = 2016,
        month = sep,
       volume = {117},
       number = {13},
          eid = {131302},
        pages = {131302},
          doi = {10.1103/PhysRevLett.117.131302},
archivePrefix = {arXiv},
       eprint = {1605.07178},
 primaryClass = {astro-ph.CO},
       adsurl = {https://ui.adsabs.harvard.edu/abs/2016PhRvL.117m1302S}
}

@ARTICLE{Campanelli2011,
       author = {{Campanelli}, Leonardo},
        title = "{Helium-4 synthesis in an anisotropic universe}",
      journal = {\prd},
         year = 2011,
        month = dec,
       volume = {84},
       number = {12},
          eid = {123521},
        pages = {123521},
          doi = {10.1103/PhysRevD.84.123521},
archivePrefix = {arXiv},
       eprint = {1112.2076},
 primaryClass = {astro-ph.CO},
       adsurl = {https://ui.adsabs.harvard.edu/abs/2011PhRvD..84l3521C}
}

@ARTICLE{Pontzen2009,
       author = {{Pontzen}, Andrew},
        title = "{Rogues' gallery: The full freedom of the Bianchi CMB anomalies}",
      journal = {\prd},
         year = 2009,
        month = may,
       volume = {79},
       number = {10},
          eid = {103518},
        pages = {103518},
          doi = {10.1103/PhysRevD.79.103518},
archivePrefix = {arXiv},
       eprint = {0901.2122},
 primaryClass = {astro-ph.CO},
       adsurl = {https://ui.adsabs.harvard.edu/abs/2009PhRvD..79j3518P}
}

@ARTICLE{Darling2018,
       author = {{Darling}, Jeremy and {Truebenbach}, Alexandra E. and {Paine}, Jennie},
        title = "{Astrometric Limits on the Stochastic Gravitational Wave Background}",
      journal = {\apj},
         year = 2018,
        month = jul,
       volume = {861},
       number = {2},
          eid = {113},
        pages = {113},
          doi = {10.3847/1538-4357/aac772},
archivePrefix = {arXiv},
       eprint = {1804.06986},
 primaryClass = {astro-ph.IM},
       adsurl = {https://ui.adsabs.harvard.edu/abs/2018ApJ...861..113D}
}

@ARTICLE{Belgacem2020,
       author = {{Belgacem}, Enis and {Kamionkowski}, Marc},
        title = "{Chirality of the gravitational-wave background and pulsar-timing arrays}",
      journal = {\prd},
         year = 2020,
        month = jul,
       volume = {102},
       number = {2},
          eid = {023004},
        pages = {023004},
          doi = {10.1103/PhysRevD.102.023004},
archivePrefix = {arXiv},
       eprint = {2004.05480},
 primaryClass = {astro-ph.CO},
       adsurl = {https://ui.adsabs.harvard.edu/abs/2020PhRvD.102b3004B}
}

@ARTICLE{Chu2021,
       author = {{Chu}, Yu-Kuang and {Liu}, Guo-Chin and {Ng}, Kin-Wang},
        title = "{Spherical harmonic analysis of anisotropies in polarized stochastic gravitational-wave background with interferometry experiments}",
      journal = {\prd},
         year = 2021,
        month = mar,
       volume = {103},
       number = {6},
          eid = {063528},
        pages = {063528},
          doi = {10.1103/PhysRevD.103.063528},
archivePrefix = {arXiv},
       eprint = {2002.01606},
 primaryClass = {gr-qc},
       adsurl = {https://ui.adsabs.harvard.edu/abs/2021PhRvD.103f3528C}
}

@ARTICLE{Orlando2021,
       author = {{Orlando}, Giorgio and {Pieroni}, Mauro and {Ricciardone}, Angelo},
        title = "{Measuring parity violation in the Stochastic Gravitational Wave Background with the LISA-Taiji network}",
      journal = {\jcap},
         year = 2021,
        month = mar,
       volume = {2021},
       number = {3},
          eid = {069},
        pages = {069},
          doi = {10.1088/1475-7516/2021/03/069},
archivePrefix = {arXiv},
       eprint = {2011.07059},
 primaryClass = {astro-ph.CO},
       adsurl = {https://ui.adsabs.harvard.edu/abs/2021JCAP...03..069O}
}

@ARTICLE{Domcke2020,
       author = {{Domcke}, Valerie and {Garc{\'\i}a-Bellido}, Juan and {Peloso}, Marco and {Pieroni}, Mauro and {Ricciardone}, Angelo and {Sorbo}, Lorenzo and {Tasinato}, Gianmassimo},
        title = "{Measuring the net circular polarization of the stochastic gravitational wave background with interferometers}",
      journal = {JCAP},
         year = 2020,
        month = may,
       volume = {2020},
       number = {5},
          eid = {028},
        pages = {028},
          doi = {10.1088/1475-7516/2020/05/028},
archivePrefix = {arXiv},
       eprint = {1910.08052},
 primaryClass = {astro-ph.CO},
       adsurl = {https://ui.adsabs.harvard.edu/abs/2020JCAP...05..028D}
}

@ARTICLE{Martinovic2021,
       author = {{Martinovic}, Katarina and {Badger}, Charles and {Sakellariadou}, Mairi and {Mandic}, Vuk},
        title = "{Searching for parity violation with the LIGO-Virgo-KAGRA network}",
      journal = {\prd},
         year = 2021,
        month = oct,
       volume = {104},
       number = {8},
          eid = {L081101},
        pages = {L081101},
          doi = {10.1103/PhysRevD.104.L081101},
archivePrefix = {arXiv},
       eprint = {2103.06718},
 primaryClass = {gr-qc},
       adsurl = {https://ui.adsabs.harvard.edu/abs/2021PhRvD.104h1101M}
}

@ARTICLE{Pen1997,
       author = {{Pen}, Ue-Li and {Seljak}, Uros and {Turok}, Neil},
        title = "{Power Spectra in Global Defect Theories of Cosmic Structure Formation}",
      journal = {\prl},
         year = 1997,
        month = sep,
       volume = {79},
       number = {9},
        pages = {1611-1614},
          doi = {10.1103/PhysRevLett.79.1611},
archivePrefix = {arXiv},
       eprint = {astro-ph/9704165},
 primaryClass = {astro-ph},
       adsurl = {https://ui.adsabs.harvard.edu/abs/1997PhRvL..79.1611P}
}

@ARTICLE{Turok1998,
       author = {{Turok}, Neil and {Pen}, Ue-Li and {Seljak}, Uro{\v{s}}},
        title = "{Scalar, vector, and tensor contributions to CMB anisotropies from cosmic defects}",
      journal = {\prd},
         year = 1998,
        month = jul,
       volume = {58},
       number = {2},
          eid = {023506},
        pages = {023506},
          doi = {10.1103/PhysRevD.58.023506},
archivePrefix = {arXiv},
       eprint = {astro-ph/9706250},
 primaryClass = {astro-ph},
       adsurl = {https://ui.adsabs.harvard.edu/abs/1998PhRvD..58b3506T}
}

@ARTICLE{Sorbo2011,
       author = {{Sorbo}, Lorenzo},
        title = "{Parity violation in the Cosmic Microwave Background from a pseudoscalar inflaton}",
      journal = {\jcap},
         year = 2011,
        month = jun,
       volume = {2011},
       number = {6},
          eid = {003},
        pages = {003},
          doi = {10.1088/1475-7516/2011/06/003},
archivePrefix = {arXiv},
       eprint = {1101.1525},
 primaryClass = {astro-ph.CO},
       adsurl = {https://ui.adsabs.harvard.edu/abs/2011JCAP...06..003S}
}

@ARTICLE{Renzini2022,
       author = {{Renzini}, Arianna I. and {Goncharov}, Boris and {Jenkins}, Alexander C. and {Meyers}, Patrick M.},
        title = "{Stochastic Gravitational-Wave Backgrounds: Current Detection Efforts and Future Prospects}",
      journal = {Galaxies},
         year = 2022,
        month = feb,
       volume = {10},
       number = {1},
          eid = {34},
        pages = {34},
          doi = {10.3390/galaxies10010034},
archivePrefix = {arXiv},
       eprint = {2202.00178},
 primaryClass = {gr-qc},
       adsurl = {https://ui.adsabs.harvard.edu/abs/2022Galax..10...34R}
}

@ARTICLE{Campeti2021,
       author = {{Campeti}, Paolo and {Komatsu}, Eiichiro and {Poletti}, Davide and {Baccigalupi}, Carlo},
        title = "{Measuring the spectrum of primordial gravitational waves with CMB, PTA and laser interferometers}",
      journal = {\jcap},
         year = 2021,
        month = jan,
       volume = {2021},
       number = {1},
          eid = {012},
        pages = {012},
          doi = {10.1088/1475-7516/2021/01/012},
archivePrefix = {arXiv},
       eprint = {2007.04241},
 primaryClass = {astro-ph.CO},
       adsurl = {https://ui.adsabs.harvard.edu/abs/2021JCAP...01..012C}
}

@ARTICLE{Vanvlasselaer2026,
       author = {{Vanvlasselaer}, Miguel},
        title = "{Inspiralling Binary merger parameter reconstruction with lunar and satellite Laser Ranging}",
      journal = {arXiv e-prints},
         year = 2026,
        month = aug,
          eid = {arXiv:2608.28753},
        pages = {arXiv:2608.28753},
          doi = {10.48550/arXiv.2608.28753},
archivePrefix = {arXiv},
       eprint = {2608.28753},
 primaryClass = {gr-qc},
       adsurl = {https://ui.adsabs.harvard.edu/abs/2026arXiv260828753V}
}

@ARTICLE{Hui2013,
       author = {{Hui}, Lam and {McWilliams}, Sean T. and {Yang}, I.-Sheng},
        title = "{Binary systems as resonance detectors for gravitational waves}",
      journal = {\prd},
         year = 2013,
        month = apr,
       volume = {87},
       number = {8},
          eid = {084009},
        pages = {084009},
          doi = {10.1103/PhysRevD.87.084009},
archivePrefix = {arXiv},
       eprint = {1212.2623},
 primaryClass = {gr-qc},
       adsurl = {https://ui.adsabs.harvard.edu/abs/2013PhRvD..87h4009H}
}

@ARTICLE{2009PhLB..678...37D,
       author = {{Dimopoulos}, Savas and {Graham}, Peter W. and {Hogan}, Jason M. and {Kasevich}, Mark A. and {Rajendran}, Surjeet},
        title = "{Gravitational wave detection with atom interferometry}",
      journal = {Physics Letters B},
         year = 2009,
        month = jul,
       volume = {678},
       number = {1},
        pages = {37-40},
          doi = {10.1016/j.physletb.2009.06.011},
archivePrefix = {arXiv},
       eprint = {0712.1250},
 primaryClass = {gr-qc},
       adsurl = {https://ui.adsabs.harvard.edu/abs/2009PhLB..678...37D}
}

@MISC{GaiaDataReleases,
    author       = {{European Space Agency}},
    title        = "{Gaia Data Releases}",
    year         = 2026,
    url          = {https://www.cosmos.esa.int/web/gaia/release},
    note         = {Accessed September 29, 2026}
}






\end{document}